%% file: draft.tex
\documentclass[a4paper,11pt]{article}
\pdfoutput=1 % if your are submitting a pdflatex (i.e. if you have
\usepackage{jheppub} % for details on the use of the package, please
\usepackage[T1]{fontenc} % if needed

\RequirePackage[displaymath]{lineno} % Display line numbers
\usepackage{epstopdf}
\usepackage{epsfig}
\usepackage{graphicx}% Include figure files
\usepackage{dcolumn}% Align table columns on decimal point
\usepackage{bm}% bold math
\usepackage{ltablex,booktabs}
\usepackage{overpic}
\usepackage{subfigure}
\usepackage{float}
\usepackage{color}
\usepackage{amsmath}
\usepackage{mathcomp}
\usepackage{mathrsfs}
\usepackage{multirow}
\usepackage{rotating}
\usepackage{amssymb}
\usepackage{gensymb}
\usepackage{amsmath}
\usepackage{tabularx}
\usepackage{threeparttable}
\usepackage{verbatim}

\usepackage{graphicx} % 加载graphicx宏包，用于插入图片
\usepackage{amsmath} % 加载amsmath宏包，用于数学公式

\newcommand{\ee}{e^+e^-}
\newcommand{\Lambdabar}{\bar{\Lambda}}
\newcommand{\Omegabar}{\bar{\Omega}^+}

\title{\boldmath Search for $\Delta S=2$ nonleptonic hyperon decays $\Omega^-\to\Sigma^{0}\pi^{-}$ and  $\Omega^-\to nK^{-}$}

\collaborationImg{\includegraphics[width=0.15\textwidth, angle=90]{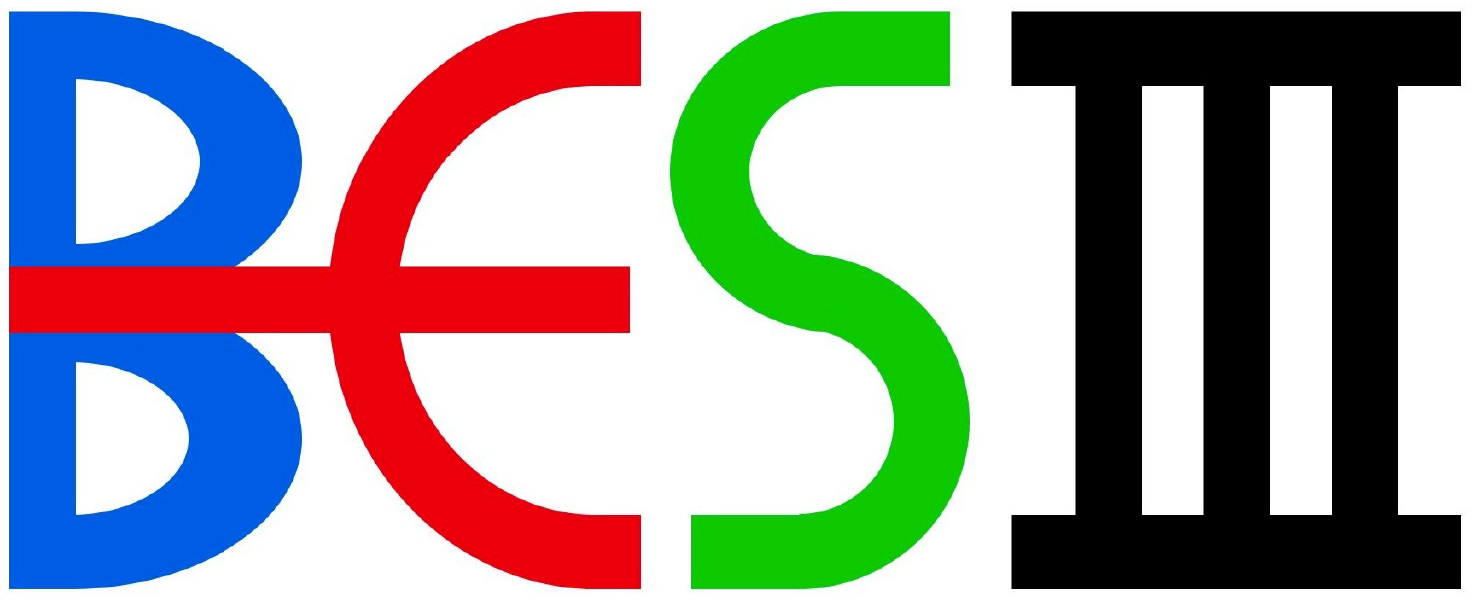}}
\collaboration{BESIII Collaboration}

\author{\input{authorlist}}

\abstract{ Using $(27.12 \pm 0.14) \times 10^{8}$ $\psi(3686)$ events
  collected by the BESIII detector at the center-of-mass energy of
  $\sqrt{s} = 3.686$ GeV, we search for the first time for two
  nonleptonic hyperon decays that change strangeness by two units,
  $\Omega^-\to\Sigma^{0}\pim$ and $\Omega^-\to nK^{-}$. No significant
  signal is observed.  The upper limits on their decay branching
  fractions are determined to be
  $\mathcal{B}(\Omega^-\to\Sigma^{0}\pim) \textless 5.4\times 10^{-4}$
  and $\mathcal{B}(\Omega^-\to nK^{-}) \textless 2.4\times 10^{-4}$ at
  the $90\%$ confidence level.  }
\arxivnumber{}

\begin{document}
\maketitle 
\flushbottom

%------------------------Introduction------------------------------------------------------
\section{Introduction}
\label{sec:introduction}

In the Standard Model (SM), nonleptonic hyperon decays involving a
change in strangeness by two units ($\Delta S = 2$) are highly
suppressed. The branching fractions (BFs) of these decays in the SM
are at the level of $10^{-17}-10^{-12}$~\cite{Theroy1}, which is far
below the existing experimental limits. Currently, there are only two $\Delta S = 2$ processes observed, $K^0-\bar{K}^0$ mixing and $B_s-\bar{B}_s$ mixing.  The $\Delta S
= 2$ nonleptonic hyperon decays may serve as probes of new
physics~\cite{Theroy1,Theroy2,Theroy3}, where the BFs could be
enhanced to the level of $10^{-10}-10^{-7}$ when beyond the SM
effects are considered~\cite{Theroy1}.  Though these
BFs may still be beyond current experimental sensitivity, it is still
worthwhile to search for these decays.

\begin{figure}[htbp]
    \begin{center}
        \mbox{
            \put(-210, 0){
                \begin{overpic}[width = 0.48\linewidth]{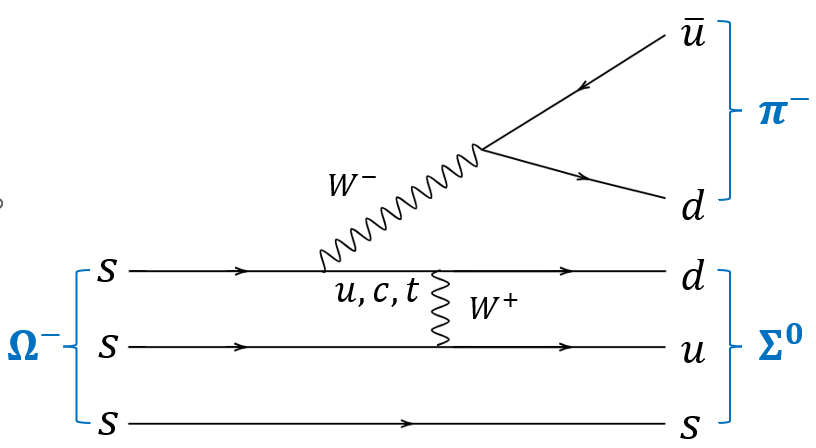}
                \end{overpic}
            }
            \put(-1, 3){\begin{overpic}[width = 0.49\linewidth]{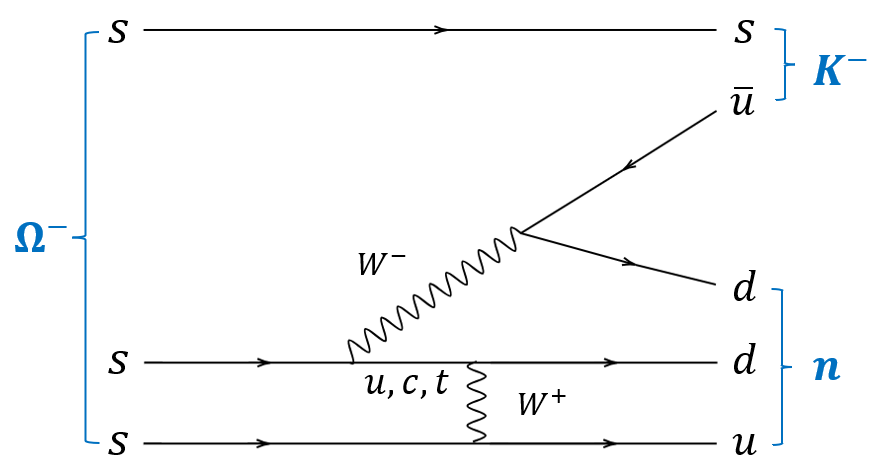}
            \end{overpic}}
            \put(-113, -5) { $\bf (a)$}
            \put(100, -5) { $\bf (b)$}
        }
    \end{center}
     \vspace{-8pt} 
    \caption{
       Feynman diagrams contributing to the $\Delta S=2$ nonleptonic hyperon decays
        (a) $\Omega^-\to\Sigma^{0}\pim$ and (b) $\Omega^-\to n K^-$  in the SM. }
    \label{fig:feyman}
\end{figure}

Until now, there have been many searches for the $\Delta S = 2$ decays
in the spin~$1/2$ hyperon sector, such as $\mathcal{B}(\Xi^0\to p
\pi^-) \textless 8.2\times 10^{-6}$~\cite{lam_ppi},
$\mathcal{B}(\Xi^0\to p e^- \bar{\nu}_e)\textless 1.3\times
10^{-3}$~\cite{Dauber:1969hg}, $\mathcal{B}(\Xi^-\to n\pi^-) \textless
1.9\times 10^{-5}$ ~\cite{Xim_npim}, $\mathcal{B}(\Xi^-\to p
\pi^-\pi^-) \textless 4.0\times 10^{-4}$~\cite{PDG}, etc., which are
all set at the 90$\%$ confidence level (C.~L.).  However, in the
spin~$3/2$ hyperon sector, only one upper limit has been set,
$\mathcal{B}(\Omega^- \to \Lambda \pi^-)<2.9\times
10^{-6}$~\cite{lam_ppi}.  The potential $\Delta S = 2$ decays,
$\Omega^-\to\Sigma^{0}\pim$ and $\Omega^-\to n K^-$, as illustrated in
Fig.~\ref{fig:feyman}, have not been explored so far. According to Ref.~\cite{Theroy1}, the decay $\Omega^-\to n K^-$ could have different new physics contributions compared to the $\Omega^- \to \Lambda \pi^-$ decay.
Therefore, it is of interest also to experimentally search for the decay $\Omega^-\to n K^-$, as well as $\Omega^-\to\Sigma^{0}\pim$.

In this paper, based on approximately $1.7\times 10^5$ $\Omega^-
\bar{\Omega}^+$ pairs~\cite{lihaibo} produced from $(27.12 \pm 0.14)
\times 10^{8}$ $\psi(3686)$ events \cite{psip} collected with the
BESIII detector in 2009, 2012 and 2021, we present the first searches
for $\Omega^- \to \Sigma^0 \pi^-$ and $\Omega^- \to n K^-$ decays (the
charge conjugated decays are always implied), using a double-tag (DT)
method. The single-tag (ST) $\bar{\Omega}^+$ hyperons are
reconstructed via the decay $\bar{\Omega}^+ \to \bar{\Lambda} K^+$.
Events where a signal candidate is reconstructed with the particles
recoiling against the ST $\bar{\Omega}^+$ hyperon are DT events.  The
BF of a signal decay is determined by \begin{equation}
    \label{eq:BFcalcu}
    \mathcal{B}_{\rm sig} = \frac{{N}_{\rm DT} \cdot \epsilon_{\rm ST}}{{N}_{\rm ST} \cdot \epsilon_{\rm DT}} = \frac{{N}_{\rm DT}}{{N}_{\rm ST} \cdot \epsilon_{\rm sig}},
\end{equation}
where ${N}_{\rm ST}$ and ${N}_{\rm DT}$ represent the ST and DT
yields, respectively, and $\epsilon_{\rm sig} = \epsilon_{\rm
  DT}/\epsilon_{\rm ST}$ is the signal efficiency in the presence of
an ST $\bar{\Omega}^+$ hyperon, where $\epsilon_{\rm ST}$ and
$\epsilon_{\rm DT}$ are the ST and DT efficiencies, respectively.

% --------------BESIII--and--MC-simulation------------------------------------------------------
\section{BESIII detector and Monte Carlo simulation}
\label{sec:BESIII and MC}

The BESIII detector~\cite{BESIII:2009fln,energyregion1} records symmetric $e^+e^-$ collisions 
provided by the BEPCII storage ring~\cite{storagering,energyregion2}
in the center-of-mass energy range from 2.0 to 4.95~GeV,
with a peak luminosity of $1.1 \times 10^{33}\;\text{cm}^{-2}\text{s}^{-1}$ 
achieved at $\sqrt{s} = 3.773\;\text{GeV}$. 
  The cylindrical core of the BESIII detector covers 93\% of the full solid angle and consists of a helium-based
 multilayer drift chamber~(MDC), a plastic scintillator time-of-flight
system~(TOF), and a CsI(Tl) electromagnetic calorimeter~(EMC),
which are all enclosed in a superconducting solenoidal magnet
providing a 1.0~T(0.9 T in 2012) magnetic field.
The solenoid is supported by an
octagonal flux-return yoke with resistive plate counter muon
identification modules interleaved with steel. 
The charged-particle momentum resolution at $1~{\rm GeV}/c$ is
$0.5\%$, and the 
$dE/dx$
resolution is $6\%$ for electrons
from Bhabha scattering. The EMC measures photon energies with a
resolution of $2.5\%$ ($5\%$) at $1$~GeV in the barrel (end cap)
region. The time resolution in the TOF barrel region is 68~ps, while
that in the end cap region was 110~ps. The end cap TOF system was upgraded in 2015 using multigap resistive plate chamber technology, providing a time
resolution of 60 ps, which benefits about 83\% of the data  used in this analysis~\cite{60ps1,60ps2,60ps3}.

Simulated data samples produced with the {\sc
  geant4}-based~\cite{geant4} Monte Carlo (MC) software~\cite{MC}
which includes the geometric description of the BESIII detector and
the detector response, are used to determine the detection efficiency
and estimate the backgrounds.  The simulation includes the beam energy
spread and initial state radiation (ISR) in the $e^+e^-$ annihilations
modeled with the generator {\sc kkmc}~\cite{kkmc}.  A sample of
simulated inclusive $\psi(3686)$ events, which includes both the
production of the $\psi(3686)$ resonance and the continuum processes
incorporated in {\sc kkmc}, is used to estimate the background events.

All particle decays are modeled with {\sc
  evtgen}~\cite{evtgen1,evtgen2} using the BFs either taken from the
PDG~\cite{PDG}, when available, or otherwise estimated with {\sc
  lundcharm}~\cite{Chen:2000tv, Yang:2014vra}.  Final state radiation
from charged final state particles is incorporated using {\sc
  photos}~\cite{photon}.  For the signal decays, three signal MC samples are
used. On the tag side, $\psi(3686) \to \Omega^- \bar{\Omega}^+$,
$\bar{\Omega}^+ \to \bar{\Lambda}(\to \bar{p}\pi^+) K^+$, which is
generated according to the angular distributions measured in
Ref.~\cite{BESIII:2020lkm}, is used to determine the ST efficiency.
 The signal decays, $\Omega^- \to$ $X$, $\Omega^-
\to \Sigma^0(\to X) \pi^-$,  and $\Omega^- \to n K^-$, are generated uniformly  in phase space.  The final  state $X$ indicates
inclusive decay. The first one is used to determine the ST efficiency, and the later two are used to determine the DT efficiencies.
 The ST sample consists of 2.54 million events. Each of the two DT samples contains 1.27 million events.

%--------------------event-selection----------------------------------------------------------
\section{Data analysis}
\label{sec:event selection}
The decay chains of interest are $\psi(3686) \to \Omega^-
\bar{\Omega}^+$, $\bar{\Omega}^+ \to \bar{\Lambda}(\to \bar{p}\pi^+)
K^+$, $\Omega^- \to \Sigma^0(\to X) \pi^-$ and $\psi(3686) \to
\Omega^- \bar{\Omega}^+$, $\bar{\Omega}^+ \to \bar{\Lambda}(\to
\bar{p}\pi^+) K^+$, $\Omega^- \to n K^-$.  The ST $\bar{\Omega}^+$
hyperons are reconstructed via the decay $\bar{\Omega}^+ \to
\bar{\Lambda} K^+$. The charged tracks in the MDC are required to have
a polar angle $\theta$ within the MDC acceptance
$|\!\cos\theta|<0.93$, where $\theta$ is defined with respect to the
$z$-axis (the symmetry axis of the MDC).  In order to perform particle
identification (PID), the specific ionization energy loss $dE/$$dx$
and the time-of-flight information are combined to form a likelihood
$\mathcal{L}(h)~(h = p, K, \pi)$ for each hadron $h$ hypothesis.
Charged tracks with $\mathcal{L}(p)>\mathcal{L}(K)$,
$\mathcal{L}(p)>\mathcal{L}(\pi)$ and $\mathcal{L}(p)>0.001$ are
identified as protons, and those with
$\mathcal{L}(K)>\mathcal{L}(\pi)$ and $\mathcal{L}(K)>0$ as kaons.
The remaining charged tracks are assigned as pions by default.

The $\Lambdabar$ candidates are reconstructed from $\bar{p} \pi^+$
pairs, which are constrained to originate from a common vertex and are
required to have an invariant mass within the range of $[1.111,
  1.121]$ GeV/$c^2$. Vertex fits are performed to the $\Lambdabar K^+$
pairs to improve the mass resolution of the $\Omegabar$ candidates. If
there are multiple $\Omegabar$ candidates, the one with the smallest
value of $|\Delta E| = |E_{\Omegabar} - E_{\rm beam}|$ is selected for
further analysis. Here, $E_{\Omegabar}$ is the energy of the
reconstructed $\Omegabar$ candidate in the $\ee$ center-of-mass system
and $E_{\rm beam}$ is the beam energy.  Additionally, the invariant
mass of the $\Lambdabar K^+$ combination~($M_{\Lambdabar K^+}$) must
be in the $\Omegabar$ signal region, defined as $[1.664, 1.680]$
GeV/$c^2$.

\begin{figure}[htbp]
    \centering
    \begin{turn}{0}
        \begin{overpic}[width=0.65\textwidth]{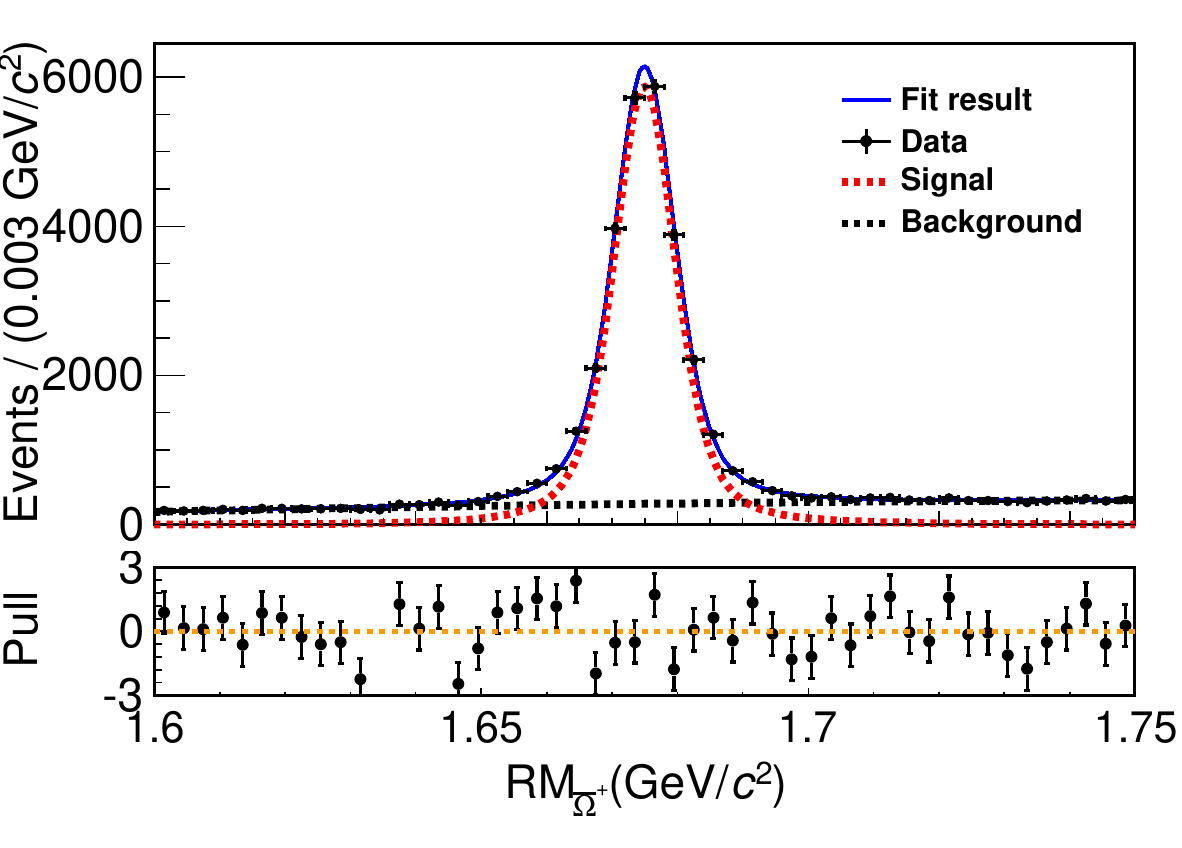}
        \end{overpic}
    \end{turn}
    \vspace{-15pt} 
    \caption{The distribution of $RM_{\Omegabar}$.  In the top panel,
      the dots with error bars are data, the blue solid line is the total fit result, and the black dashed and red dashed lines represent the fitted background and signal shapes, respectively. 
      The bottom panel displays the pull
      distribution, illustrating the residuals between the data and
      the fitted model and normalized by their uncertainties.}
    \label{fig:SingleTag}
\end{figure}

To determine the ST yield, a fit is applied to the recoil-mass
spectrum against the reconstructed $\Omegabar$ ($RM_{\Omegabar}$), as
shown in Fig.~\ref{fig:SingleTag}.  In the fit, the signal shape is
described by the MC simulated shape convolved with a Gaussian function
with free parameters, where the Gaussian function is used to account
for the difference in mass resolution between data and MC simulation.
The background shape is described by a second-order Chebychev
polynomial.  For the search for $\Omega^-\to\Sigma^{0}\pim$ and
$\Omega^-\to n K^-$, the signal region of $RM_{\Omegabar}$ is defined
as $[1.652, 1.695]$~GeV/$c^2$. The number of ST $\Omegabar$ hyperons
in the signal region is determined to be $25819 \pm 188$, and the ST
efficiency is estimated to be 21.11\% based on MC simulation.  Events
in $\Omegabar$ sideband region, defined as $M_{\Lambdabar K^+} \in
[1.648, 1.656]$ $\cup$ $[1.688, 1.696]$~GeV/$c^2$, are used to study
the backgrounds in the $RM_{\Omegabar}$ signal region, and we find
that the background distribution is smooth with no peaking background.

%\section{Signal side event selection for $\Omega^- \to \Sigma^0 \pi^-$} 
%\label{1} 
Candidates for $\Omega^- \to \Sigma^0 \pi^-$ are
selected from the surviving tracks in the system recoiling against the
ST $\bar{\Omega}^+$ hyperons. To improve the DT efficiency, only the
bachelor $\pi^-$ (from $\Omega^-$ decay) is reconstructed in the
signal side. The polar angle of the bachelor $\pi^-$ must satisfy
$|\!\cos\theta|<0.93$.  The $\pi^-$ candidates are identified using
information measured by the MDC ($dE/dx$), TOF, and EMC. The
probabilities for the pion and kaon hypotheses are calculated, and the
pion candidate is required to satisfy
$\mathcal{L}(\pi)>\mathcal{L}(K)$ and $\mathcal{L}(\pi) >0 $. If there
is more than one $\pi^-$ candidate, the one with the highest momentum
is retained.

For the decay $\Omega^-\to nK ^-$, a similar approach is used, where
only the bachelor $K^-$ (from $\Omega^-$ decay) is reconstructed on
the signal side. The polar angle of the bachelor $K^-$ must satisfy
$|\!\cos\theta|<0.93$.  The $K^-$ candidates are identified using the
information measured by the MDC ($dE/dx$), TOF, and EMC. The
probabilities for the kaon and pion hypotheses are calculated, and the
kaon candidate is required to satisfy
$\mathcal{L}(K)>\mathcal{L}(\pi)$ and $\ \mathcal{L}(K) >0 $. Additionally,  the number of charged tracks ($N_{tracks}$) including the
single and tag sides is required to be four to further suppress
background events.

\begin{figure}[htbp]
	\centering
	\begin{overpic}[width=0.49\textwidth]{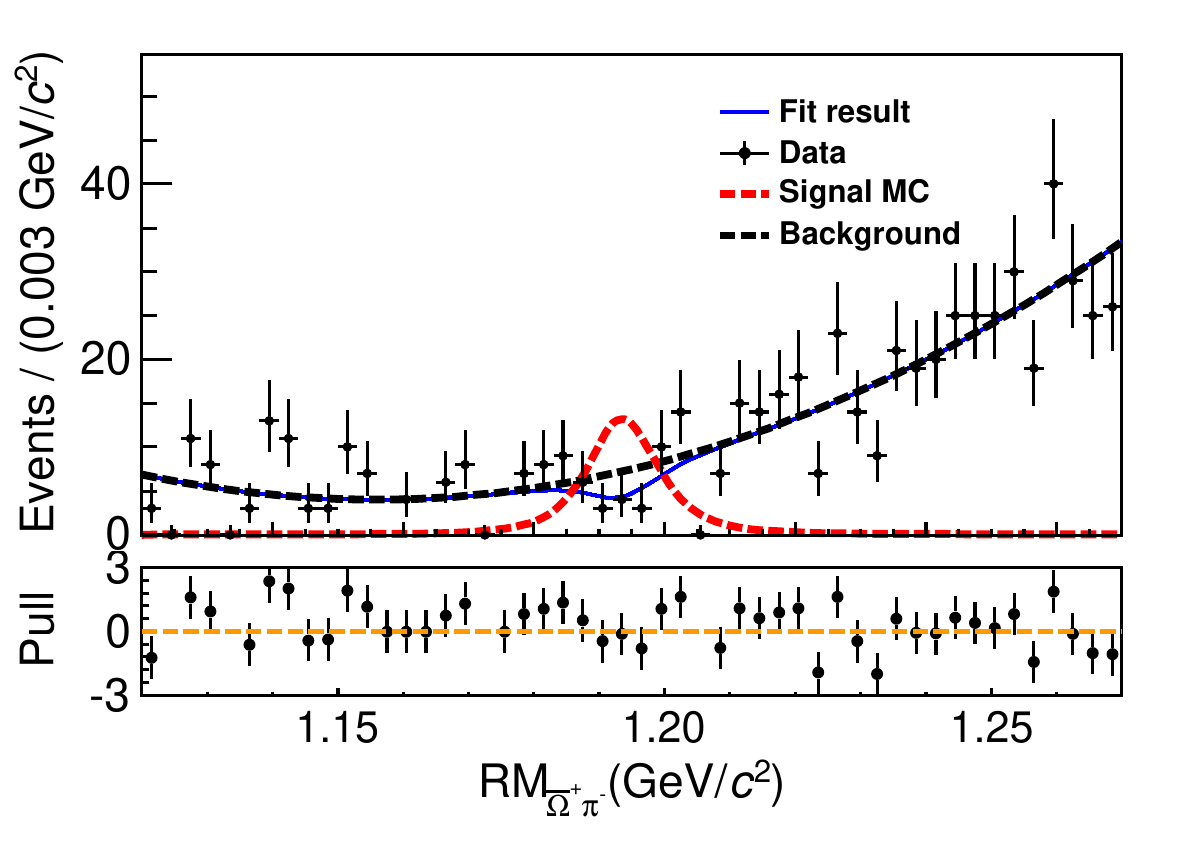}
		\put(18,60){$\bf (a)$}
	\end{overpic}
	\begin{overpic}[width=0.49\textwidth]{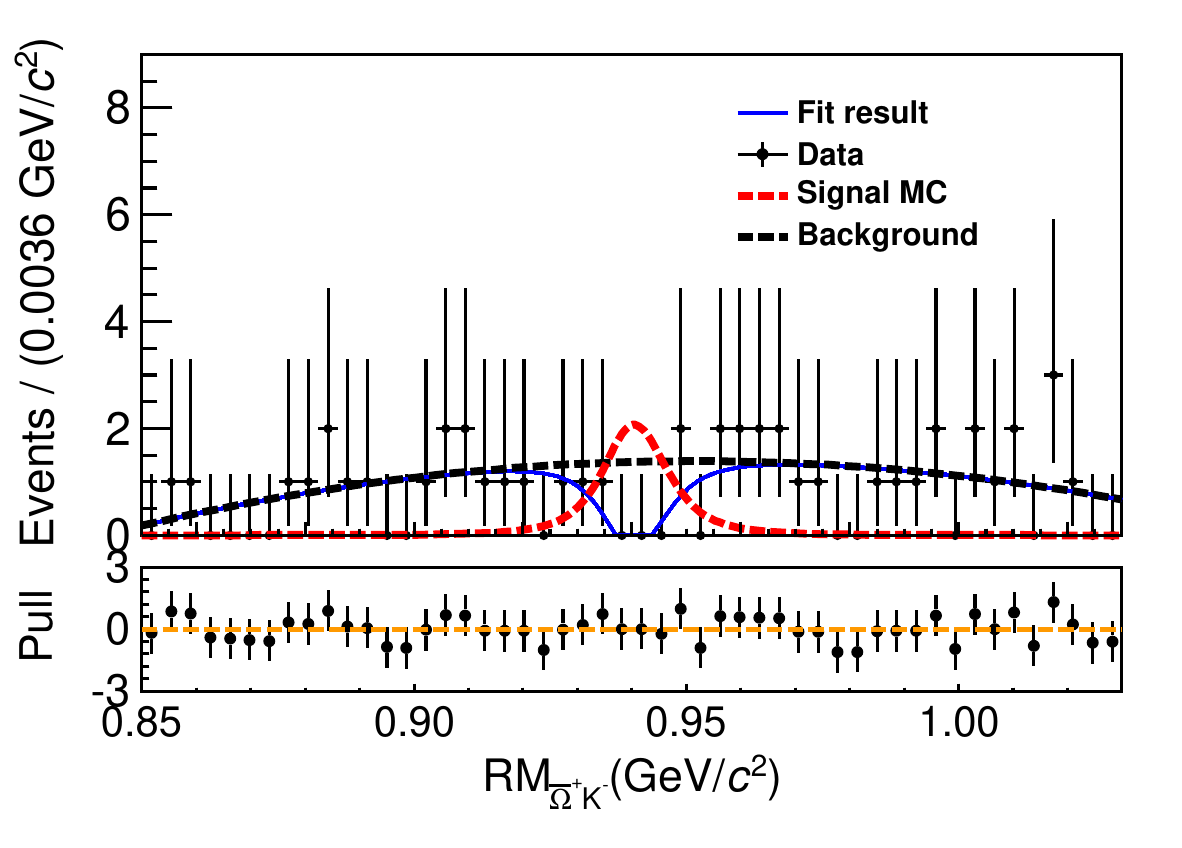}
		\put(18,60){$\bf (b)$}
	\end{overpic}
	 \vspace{-13pt} 
\caption{The distributions of (a) $RM_{\bar{\Omega}^+ \pi^-}$
  and (b) $RM_{\bar{\Omega}^+ K^-}$. In the top panel, the dots with error bars are data, the blue solid and black dashed lines are the total fit result and the fitted background shape, respectively. The red dashed line is the signal shape obtained from the signal MC with arbitrary normalization. 
  The bottom panels display the pull distributions, illustrating the residuals between the data and the fitted model normalized by their uncertainties.}
	\label{fig:fit}
\end{figure}

The recoiling mass distribution of $\bar{\Omega}^+ h$ ($RM_{\bar{\Omega}^+ h}$) 
is used to extract the DT yield,  where $h=\pi$ or $K$.
Before that, potential backgrounds in the studies of  $\Omega^- \to \Sigma^0 \pi^-$ and $\Omega^- \to n K^-$ are investigated by analyzing the inclusive MC
sample and the events in the $M_{\Lambdabar K^+}$ and $RM_{\Omegabar}$
sideband regions from data. The sideband regions in $RM_{\bar{\Omega}^+}$ are defined as $[1.718, 1.739]$$\cup$  $[1.608, 1.630]$~GeV/$c^2$. For each decay, the background shape is found to be smooth in the $RM_{\bar{\Omega}^+ h}$ spectrum. 
Then the number of DT events in data is obtained by fitting the $\bar{\Omega}^+ h$ distribution. 
In the fit, the signal shape is described by the
simulated shape derived from signal MC sample, and the background
shape is described with a second-order Chebychev polynomial. The fit
result is shown in Fig.~\ref{fig:fit}. The numbers of DT events are
$-15.4^{+10.0 }_{-9.1}$ for $\Omega^- \to \Sigma^0\pi^-$ and $-8.3^{+
  5.5}_{-3.7}$ for $\Omega^-\to nK^-$. Since no significant signal is
observed for each decay, we set upper limits on the BFs of these two
decays.
 
% Potential backgrounds in the studies of  $\Omega^- \to \Sigma^0 \pi^-$ and $\Omega^- \to n K^-$ are investigated by analyzing the inclusive MC
%sample and the events in the $M_{\Lambdabar K^+}$ and $RM_{\Omegabar}$
%sideband regions from data. The sideband regions in $RM_{\bar{\Omega}^+}$ are defined as %$[1.718, 1.739]$$\cup$  $[1.608, 1.630]$~GeV/$c^2$. For each decay, the background shapes are %found to be flat in the signal region of $\Sigma^0$ or $n$.

%The number of DT events in data is obtained by fitting the recoiling
%mass distribution of $\bar{\Omega}^+ h$ ($RM_{\bar{\Omega}^+ h}$), where
%$h=\pi$ or $K$. In the fit, the signal shape is described by the
%simulated shape derived from signal MC sample, and the background
%shape is described with a second-order Chebychev polynomial. The fit
%result is shown in Fig.~\ref{fig:fit}. The numbers of DT events are
%$-15.4^{+10.0 }_{-9.1}$ for $\Omega^- \to \Sigma^0\pi^-$ and $-8.3^{+
%  5.5}_{-3.7}$ for $\Omega^-\to nK^-$. Since no significant signal is
%observed for each decay, we set upper limits on the BFs of these two
%decays.

%--------------------background-study----------------------------------------------------------
%\section{Signal side event selection for $\Omega^- \to n K^-$}
%\label{sec:2}

\section{Systematic uncertainty}

The sources of systematic uncertainties are classified into two
categories, which are multiplicative and additive. The
multiplicative systematics affect the efficiency while the additive
ones affect the signal yield.

The multiplicative systematic uncertainties mainly stem from the
imperfect modeling of data in our simulation. Considering the
differences between MC simulation and data, the following sources are
taken into account. The tracking and PID efficiencies of charged pions
and kaons are investigated using the control samples of $J/\psi\to
p\bar{p}\pi^{+}\pi^{-}$ and $J/\psi\to K^0_S K^{\pm}\pi^{\mp}$
\cite{MDC1,MDC2,BESIII:2023drj}, respectively. The differences in tracking and PID
efficiencies between data and MC simulation are both $1.0\%$ per
charged pion on the signal side, and the uncertainties of the tracking
and PID are both $1.0\%$ per kaon on the signal side. The systematic
uncertainty due to the signal model is studied by the
control samples of $\Omega^-\to\Xi^0\pi^-$ and $\Omega^-\to\Lambda
K^-$~\cite{hongfei}.  The differences between the efficiencies
obtained from phase space MC samples and data driven samples (re-weighted according to $\pi^-/K^-$ transverse momentum distribution in data) are taken
as the systematic uncertainties, which are $2.2\%$ and $6.9\%$ for
$\Omega^- \to \Sigma^0 \pi^-$ and $\Omega^- \to n K^-$,
respectively. The uncertainty from the $RM_{\bar{\Omega}^+}$ signal
shape for ST is estimated by an alternative signal shape obtained from
a Breit-Wigner convolved with a double Gaussian function with all free
parameters.  The resultant difference in the ST yield, 2.0$\%$, is
taken as the systematic uncertainty.  Futhermore, the uncertainty due
to the $RM_{\bar \Omega^+}$ background shape for ST is studied by
changing the second-order Chebychev polynomial to a first-order and a
third-order Chebychev polynomial. The larger difference in the ST
yield, 0.7\%, is considered as the corresponding uncertainty. The
uncertainty due to the MC statistics is assigned by using the formula
$\frac{1}{\sqrt{N}}\sqrt{\frac{1-\epsilon}{\epsilon}}$, where
$\epsilon$ is the DT efficiency and $N$ is the number of the
generated signal MC events. The systematic uncertainties are both
0.2\% for $\Omega^- \to \Sigma^0 \pi^-$ and $\Omega^- \to n K^-$. The
statistical fluctuation of the ST $\bar \Omega^+$ hyperons, 0.7\%, is
taken as a systematic uncertainty from the ST yield estimation. The
systematic uncertainty due to the signal angular distribution is
studied by the control samples of $\Omega^-\to\Xi^0\pi^-$ and
$\Omega^-\to\Lambda K^-$~\cite{hongfei}. In the search for $\Omega^-
\to n K^-$, the systematic uncertainty of the requirement on the
number of charged tracks is studied by using a control sample of
$\psi(3686)\to\Omega^{-}\bar{\Omega}^{+}$, $\bar{\Omega}^+ \to
\bar{\Lambda}(\to \bar{p}\pi^+) K^+$, $\Omega^- \to\Lambda(\to p
\pi^-) K^-$. This control sample, comprising six charged tracks,
is used to evaluate the efficiencies of the $N_{tracks} =6$ requirement
for data and MC. The efficiency difference between data and MC in the
control sample, 3.1\%, is taken as the systematic uncertainty.  All
the multiplicative uncertainties are summarized in
Table~\ref{tab:mult}. The individual uncertainties are assumed to be
independent and are combined in quadrature to obtain the total
multiplicative systematic uncertainty. Note that the systematic uncertainty of the measurement of $\Omega^- \to n K^-$ is dominated by the signal model because the kaon reconstruction efficiency is heavily dependent on its transverse momentum.

There are two sources in the additive systematic uncertainties. One
systematic uncertainty due to the $RM_{\bar{\Omega}^+h}$ signal shape
is studied by changing the MC simulated shape to double Johnson
functions~
\cite{Johnson} sharing the same mean and width parameters
determined from the fit. The tail parameters and fractions of
each signal component are fixed to values obtained from a
fit to simulated events. Another systematic uncertainty due to the
$RM_{\bar{\Omega}^+h}$ background shape is studied by changing the Chebychev
polynomial from second to third order. The results are used to
estimate the upper limits , described in the next section. 

\begin{table}[!htbp]
    \centering
    \caption{\small Multiplicative systematic uncertainties~($\%$), where the dash (-) indicates that a systematic effect is not applicable.}
    \label{tab:mult}
    \setlength{\tabcolsep}{11mm}
    \resizebox{\linewidth}{!}{
        \begin{tabular}{l|c|c}
            \hline
            \hline
            Source							&  $\Omega^- \to \Sigma^0 \pi^-$  	&  $\Omega^- \to n K^-$	 \\
            \hline
            Tracking 	                & 1.0     & 1.0      \\          
            PID 			                & 1.0     & 1.0   \\
            %	Photon selection                & 1.0     & 1.0    \\
            ST signal shape 			        & 2.0       & 2.0  \\
            ST background shape                & 0.7      & 0.7  \\
            MC statistics                   &0.2       &0.2  \\
            ST yields                       &0.7     &0.7   \\
            Signal model  &2.2  &6.9  \\
            Number of charged tracks   &-&3.1  \\
            \hline
            Total							&  3.4  &8.0 \\
            \hline
            \hline
        \end{tabular}
    }
\end{table}

\section{The upper limits on the BFs}

\begin{figure}[htbp]
	\begin{center}
        \mbox{
            \put(-220, 10){
                \begin{overpic}[width = 0.55\linewidth]{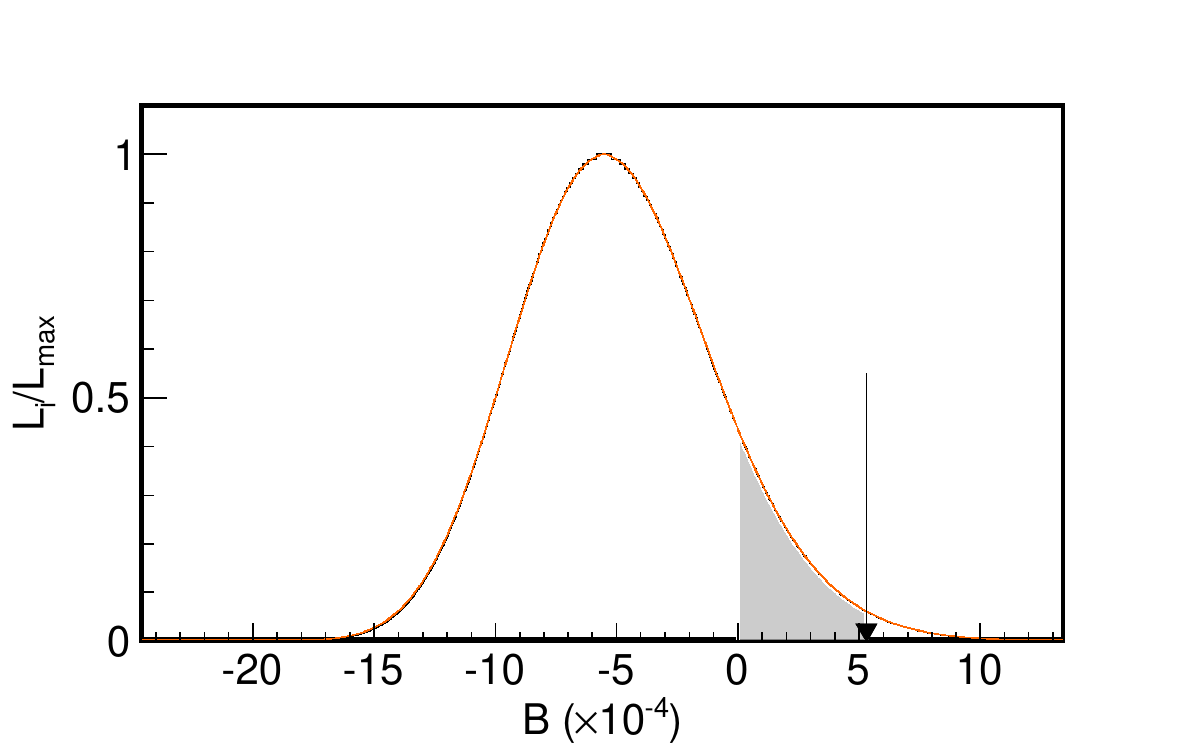}
                \end{overpic}}
            \put(-4, 10){
                \begin{overpic}[width = 0.55\linewidth]{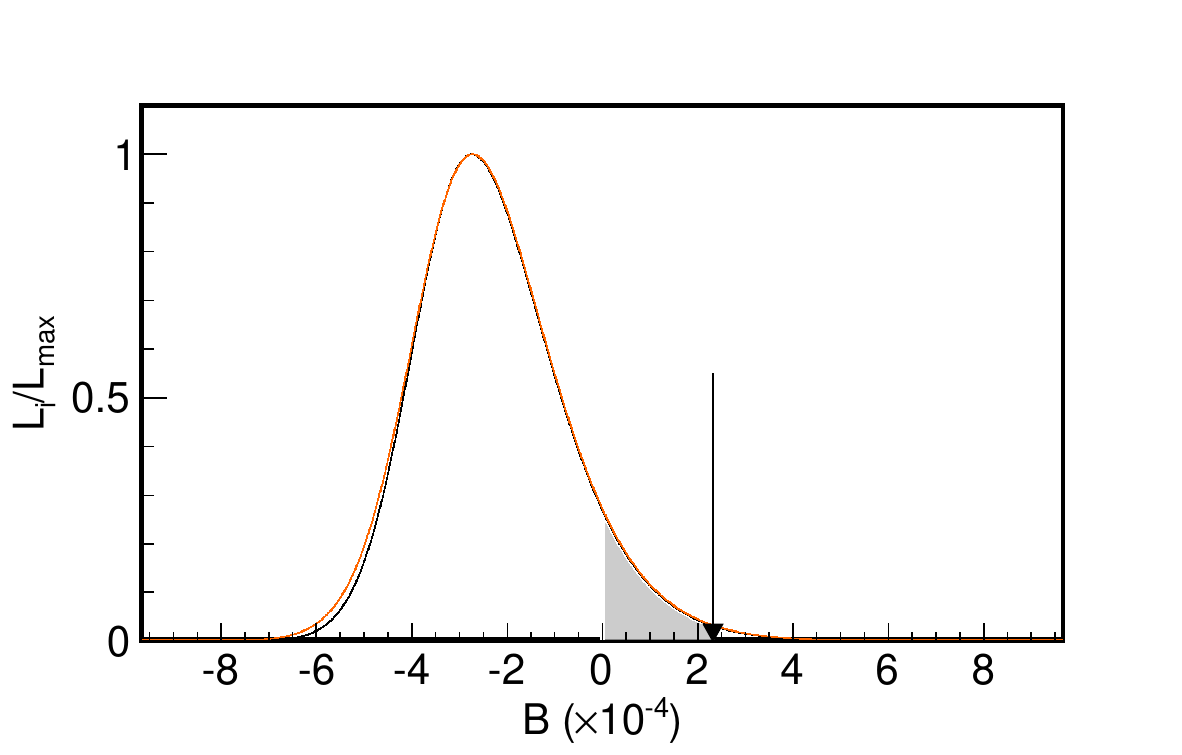}
                \end{overpic}}
      \put(-180, 125) { \bf(a)}
      \put(38, 125)   { \bf(b)}
        }
	\end{center}
	
 \vspace{-30pt} % 减小图与caption之间的距离
	\caption{The normalized likelihood distributions for (a)
	$\Omega^-\to \Sigma^0\pi^-$ and (b) $\Omega^- \to n K^-$. The
	orange curves are obtained with incorporating both
	multiplicative and additive systematic uncertainties. The
	black curves are the likelihood distibution without
	multiplicative systematic uncertainties. The arrows point the
	upper limit on the BFs at the $90\%$ C. L.}
	\label{fig:smear}
\end{figure}

The upper limits on the signal yields at the 90\% C.~L. are
determined by a Bayesian method~\cite{PhysRevD.57.3873}.  The additive uncertainties are accounted for by extracting the likelihood distributions,  and the signal shapes corresponding to the maximum upper limits among all additive items are chosen for $\Omega^- \to \Sigma^0 \pi^-$ and $\Omega^- \to n
 K^-$, respectively.
The upper limits based on these likelihood distributions and
incorporating the multiplicative systematic uncertainties in the
calculation are obtained by smearing the likelihood distribution by a
Gaussian function with a mean of zero and a width equal to
$\sigma_{\epsilon}$ as described in Ref.~\cite{UL1,UL2} with the
following formula,
\begin{equation}
     L'(n) \propto \int_{0}^{1} L(n\frac{\epsilon}{\epsilon_0})
     {\rm exp}[\frac{-(\epsilon-\epsilon_0)^2}{2\sigma^2_\epsilon}]d\epsilon,
     \label{eq:Ln} \end{equation}
 where $L(n)$ is the likelihood distribution as a function of the yield $n$.  $\epsilon_{0}$ is the signal efficiency
 and $\sigma_{\epsilon}$ is the multiplicative systematic
 uncertainty. The signal yield at the 90\% C. L., $N^{U.L.}$, is
 obtained by integrating out to 90\% of its physical region,
 $\frac{\int_0^{N^{U.L.}} L^{\prime}\left(n \right) d
 n}{\int_0^{\infty} L^{\prime}\left(n \right) d n }=0.9$. Figures
 \ref{fig:smear}(a) and \ref{fig:smear}(b) show the likelihood
 distributions incorporating the systematic uncertainties for
 $\Omega^- \to \Sigma^0 \pi^-$ and $\Omega^- \to n K^-$,
 respectively. The upper limits on the signal yields ($N^{\rm
 U.L.}_{\rm DT}$) of $\Omega^- \to \Sigma^0 \pi^-$ and $\Omega^- \to n
 K^-$ at the 90\% C. L. are 12 and 5, respectively. With the ST
 yields, and the ST and DT efficiencies obtained from the MC
 simulation, the upper limits on the BFs on the signal decays at the
 90\% C. L. are calculated by

\begin{equation}
    {\mathcal B}^{\rm U.L.}_{\rm sig} = \frac{N^{\rm U.L.}_{\rm DT}/\epsilon_{\rm DT}}{N_{\rm ST}/\epsilon_{\rm ST}}.
\label{eq:brul}
\end{equation}
 
\noindent The numerical results are shown in Table~\ref{tab:BFs}.

\begin{table}[htbp]
    \caption{The ST yield~($N_{\rm ST}$), ST efficiency ($\epsilon_{\rm ST}$), DT yields ($N^{\rm U.L.}_{\rm DT}$
    ), DT efficiencies ($\epsilon_{\rm DT}$) and the upper limits on the BFs ($\mathcal B^{\rm U.L.}$).}
    \label{tab:BFs}
    \setlength{\extrarowheight}{1.2ex}
    \renewcommand{\arraystretch}{1.0}
    \begin{center}
        \scalebox{0.82}{
            \begin{tabular} {l | c | c | c | c | c}
                \hline \hline
                Decay mode  & $N_{\rm ST}$ & $\epsilon_{\rm ST}~(\%)$ & $N^{\rm U.L.}_{\rm DT}$ & $\epsilon_{\rm DT}~(\%)$ & ${\mathcal B}^{\rm U.L.}$ ($\times 10^{-4}$)    \\
                \hline
                $\Omega^- \to \Sigma^0 \pi^-$ & \multirow{2}{2.8cm}{\centering $25819\pm188$} & \multirow{2}{1.1cm}{\centering $21.11$} & $12$ & $18.29$ & $5.4$\\
                \cline{1-1} \cline{4-6}
                $\Omega^- \to n K^-$ & &  & $5$ & $16.92$ & $2.4$\\
                \hline \hline
            \end{tabular}
        }
    \end{center}
\end{table}

%------------------------Summary------------------------------------------------------
\section{Summary}
\label{sec:summary}

Based on the sample of $(27.12 \pm 0.14) \times 10^{8}$ $\psi(3686)$
events collected by the BESIII detector, we search for the $\Delta
S=2$ decays of $\Omega^-\to\Sigma^{0}\pim$ and $\Omega^-\to nK^{-}$
for the first time.  No signal is observed. The upper limits on their
decay BFs are determined to be $\mathcal{B}(\Omega^-\to\Sigma^{0}\pim)
\textless 5.4\times 10^{-4}$ and $\mathcal{B}(\Omega^-\to nK^{-})
\textless 2.4\times 10^{-4}$ at the $90\%$ C. L.  These results are
consistent with the predictions of the SM~\cite{Theroy1} and can help
to constrain new physics models.  In the future, the proposed Super
Tau-Charm Factories~\cite{Charm-TauFactory:2013cnj,Achasov:2023gey}
have the potential to improve on the upper limits on the BFs of these
decays by at least two orders of magnitude.

\include{acknowledgement.tex}

\bibliographystyle{JHEP}
\input{draft.bbl}
% \bibliography{References.bib}

\end{document}

%% file: authorlist.tex
%% Saved at => 2023-12-15
M.~Ablikim$^{1}$, M.~N.~Achasov$^{4,c}$, P.~Adlarson$^{75}$, O.~Afedulidis$^{3}$, X.~C.~Ai$^{80}$, R.~Aliberti$^{35}$, A.~Amoroso$^{74A,74C}$, Q.~An$^{71,58,a}$, Y.~Bai$^{57}$, O.~Bakina$^{36}$
% ...
\\
\vspace{0.2cm}
(BESIII Collaboration)\\
\vspace{0.2cm} {\it
$^{1}$ Institute of High Energy Physics, Beijing 100049, People's Republic of China\\
% ...
$^{3}$ Bochum  Ruhr-University, D-44780 Bochum, Germany\\
$^{4}$ Budker Institute of Nuclear Physics SB RAS (BINP), Novosibirsk 630090, Russia\\
% ...
$^{35}$ Johannes Gutenberg University of Mainz, Johann-Joachim-Becher-Weg 45, D-55099 Mainz, Germany\\
$^{36}$ Joint Institute for Nuclear Research, 141980 Dubna, Moscow region, Russia\\
% ...
$^{57}$ Southeast University, Nanjing 211100, People's Republic of China\\
$^{58}$ State Key Laboratory of Particle Detection and Electronics, Beijing 100049, Hefei 230026, People's Republic of China\\
% ...
$^{71}$ University of Science and Technology of China, Hefei 230026, People's Republic of China\\
% ...
$^{74}$ University of Turin and INFN, (A)University of Turin, I-10125, Turin, Italy; (B)University of Eastern Piedmont, I-15121, Alessandria, Italy; (C)INFN, I-10125, Turin, Italy\\
$^{75}$ Uppsala University, Box 516, SE-75120 Uppsala, Sweden\\
% ...
$^{80}$ Zhengzhou University, Zhengzhou 450001, People's Republic of China\\

\vspace{0.2cm}
$^{a}$ Deceased\\
% ...
$^{c}$ Also at the Novosibirsk State University, Novosibirsk, 630090, Russia\\
% ...

}
%% ends here %%

%% file: acknowledgement.tex
%% Saved at => 2023-12-15
\textbf{Acknowledgement}

The BESIII Collaboration thanks the staff of BEPCII and the IHEP computing center for their strong support. The authors would like to extend thanks to Prof. Jusak Tandean for useful discussion and helpful advice. This work is supported in part by National Key R\&D Program of China under Contracts Nos. 2020YFA0406300, 2020YFA0406400; National Natural Science Foundation of China (NSFC) under Contracts Nos. 11635010, 11735014, 11835012, 11935015, 11935016, 11935018, 11961141012, 12025502, 12035009, 12035013,  12061131003, 12192260, 12192261, 12192262, 12192263, 12192264, 12192265, 12221005, 12225509, 12235017, 12342502; the Chinese Academy of Sciences (CAS) Large-Scale Scientific Facility Program; the CAS Center for Excellence in Particle Physics (CCEPP); Joint Large-Scale Scientific Facility Funds of the NSFC and CAS under Contract No. U1832207; CAS Key Research Program of Frontier Sciences under Contracts Nos. QYZDJ-SSW-SLH003, QYZDJ-SSW-SLH040; 100 Talents Program of CAS; The Institute of Nuclear and Particle Physics (INPAC) and Shanghai Key Laboratory for Particle Physics and Cosmology; European Union's Horizon 2020 research and innovation programme under Marie Sklodowska-Curie grant agreement under Contract No. 894790; German Research Foundation DFG under Contracts Nos. 455635585, Collaborative Research Center CRC 1044, FOR5327, GRK 2149; Istituto Nazionale di Fisica Nucleare, Italy; Ministry of Development of Turkey under Contract No. DPT2006K-120470; National Research Foundation of Korea under Contract No. NRF-2022R1A2C1092335; National Science and Technology fund of Mongolia; National Science Research and Innovation Fund (NSRF) via the Program Management Unit for Human Resources \& Institutional Development, Research and Innovation of Thailand under Contract No. B16F640076; Polish National Science Centre under Contract No. 2019/35/O/ST2/02907; The Swedish Research Council; U. S. Department of Energy under Contract No. DE-FG02-05ER41374.

%% ends here %%

%% file: draft.bbl
\providecommand{\href}[2]{#2}\begingroup\raggedright\endgroup